\documentclass[twocolumn,showpacs,preprintnumbers,amsmath,amssymb,floatfix,aps]{revtex4}

\usepackage{graphicx}
\usepackage{epstopdf}
\usepackage{dcolumn}
\usepackage{bm}
\usepackage{cases}

\begin{document}

\title{Tight-Binding Model for Adatoms on Graphene: Analytical Density of States, Spectral Function, and Induced Magnetic Moment}
\author{N. A.  Pike and D. Stroud}
\affiliation{Department of Physics, The Ohio State University, Columbus, OH 43210}

\date{\today}

\begin{abstract}
In the limit of low adatom concentration, we obtain exact analytic expressions for the local and total density of states (LDOS, TDOS) for a tight-binding model of adatoms on graphene.  The model is not limited to nearest-neighbor hopping but can include hopping between carbon atoms at any separation.   We also find an analytical expression for the spectral function $A({\bf k}, E)$ of an electron of Bloch vector ${\bf k}$ and energy E on the graphene lattice, to first order in the adatom concentration.   We treat the  electron-electron interaction by including a Hubbard term on the adatom, which we solve within a mean-field approximation.  For finite Hubbard $U$, we find the spin-polarized LDOS, TDOS, and spectral function self-consistently. For any choice of parameters of the tight-binding model within mean field theory,  we find a critical value of $U$ above which a moment develops on the adatom.   For most choices of parameters, we find a substantial charge transfer from the adatom to the graphene host. 
\end{abstract}


\maketitle

\section{Introduction}

Graphene is a well known allotrope of carbon in which the carbons bond in a planar $sp^2$ configuration\cite{wallace, novoselov2004}.  As a result, the single graphene sheet is effectively two-dimensional\cite{neto2009,saito1998}.   Of the four $n=2$ electrons which occupy the outer shell of a carbon atom, three are in $sp^2$ orbitals and form in-plane $\pi$ bonds between the nearest-neighbor carbon atoms, while the fourth occupies a $2p_z$ orbital.  These $2p _z$ orbitals form a band of states which is responsible for many of the characteristic electronic properties of graphene\cite{neto2009}.  Among these properties are a zero band gap at the so-called Dirac point, an electronic dispersion relation that, near the Dirac point, is equivalent to that of massless Dirac fermions,  and spin-orbit coupling  which is believed to be  small because of the low atomic number of carbon\cite{neto2009,saito1998}.  Graphene has a vast number of  potential applications, including photo-voltaic cells\cite{wang2008}, ultracapacitors\cite{liu2011, stoller2008}, and spin-transport  electronics\cite{yazyev2008,tombros,swartz}.

Recently, a number of researchers have carried out experimental and theoretical investigations into the effects of adatoms and  impurities on both the band structure and  localized magnetic moments in graphene. Among these are theoretical studies of carbon vacancies in graphene\cite{Lehtinen2004, Skrypnyk, Yazyev2007}, hydrogen atoms on the surface of graphene\cite{sofo2012}, and several other types of disorder in graphene\cite{pereira2008,wehling}.    In several of these cases and in other work\cite{wehling,Yuan2010,Rakhmanov12}, impurity effects have been treated using a tight-binding model for the electronic structure of graphene and impurities, vacancies, or adatoms.    These calculations have, however, either been carried out numerically, or in the limit of energies close to the Dirac point, where the graphene density of states can be approximated as linear\cite{wehling1}. 

Density-functional calculations for adatoms on graphene have also been carried out.  They have shown that the introduction of an adatom bonded to the surface of graphene can lead to a quasi-localized state with an energy near the Fermi energy and that the wave function of this quasi-localized state includes contributions from the orbitals of neighboring carbon atoms\cite{sofo2012, Yazyev2007}. In some cases it has been found that, even if the introduced defect or adatom is non-magnetic, a localized magnetic moment can form at the defect site\cite{Yazyev2007}.

In this paper, we extend the previous work on graphene with adatoms in two ways.   First, we show that the tight-binding model for adatoms on graphene can, in the limit of low concentrations, be solved analytically in the absence of electron-electron interactions.   Specifically, we obtain analytical expressions for the local density of states (LDOS) on the adatom, the total density of states (TDOS) of the adatom-graphene system, and the spectral function $A({\bf k}, E)$ for an electron with Bloch vector ${\bf k}$ and energy $E$ in graphene in the presence of the adatom\cite{Bena2009}.  All these results are expressed as a function of the graphene density of states, which itself is known analytically for a nearest-neighbor tight binding model\cite{Yuan2010,Hobson1953}.

Secondly, we calculate the magnetic properties of the system using the Hubbard model for the electron-electron interaction,  which we treat using a standard mean-field approximation.  This treatment leads to a transition between a non-magnetic and magnetic state above a critical value of $U$ which depends on the parameters of the tight-binding model.   In the presence of a finite $U$ our model is basically a special case of the well-known Anderson model\cite{anderson1961}, but with a linear rather than a constant density of states near the Fermi energy.   Our results include not only the magnetic moment on the adatom, but also that on the graphene sheet and the charge transfer from the adatom to the sheet, all as functions of the model parameters.

The remainder of the article is arranged as follows:  In Sec.\ II, we describe the model tight-binding Hamiltonian for the graphene-adatom system.  We also describe its generalization to include electron-electron interaction on the adatom via a Hubbard $U$ term and the mean-field treatment of this term.   In Sec.\ III, we describe the Green's function method used to analytically calculate the local density of states on the adatom, the total density of states on the graphene lattice in the presence of the adatom, and the spectral function.  In Sec.\ IV, we present numerical results for the local densities of states, total density of states, and the spectral function for the graphene-adatom system.   We also give the local densities of states in the presence of a finite Hubbard $U$ within mean-field theory, and give the magnetic moment induced by the adatom, as well as the charge transfer from the adatom to the graphene, as a function of the adatom parameters.   In Sec.\ V, we give a concluding discussion.

\section{The Model Hamiltonian}\label{model} 
Graphene is composed of two inter-penetrating triangular lattices, which we will label $\alpha$ and $\beta$, and thus two carbon atoms per primitive cell.   In the present work, we are interested in a system consisting of a perfect lattice of graphene plus a single adatom, which we will assume has one atomic orbital.   We also assume that the adatom lies at the so-called $T$ site, above one of the carbon atoms.   It has been found, using {\it ab initio} electronic structure calculations, that a several species of adatoms, including hydrogen, fluorine, and gold, do occupy a location above one of the carbon atoms\cite{nakada,chan,ishii}.

\begin{figure}[ht]
  \centering
    \includegraphics[width=0.45\textwidth]{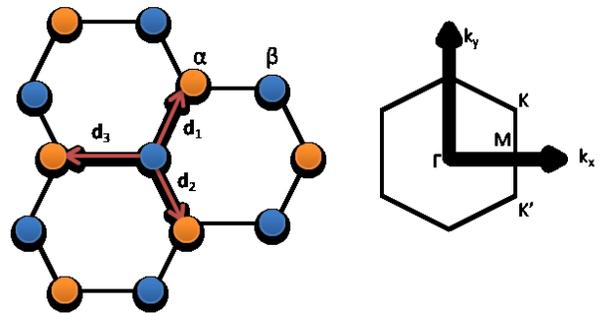}
  \caption{(Color Online) Left: graphene crystal structure showing the two interpenetrating lattices labeled as $\alpha$ and $\beta$ and the set of nearest neighbor vectors  ${\bf d}_1$, ${\bf d}_2$, and ${\bf d}_3$.  
Right: first Brillouin zone for graphene showing the high symmetry points.  This figure is a modified version of one shown in Ref.\ \cite{neto2009}. }
\label{fig:symmetry_points}
\end{figure}

We treat the graphene-adatom system using a tight-binding Hamiltonian that can, in principle, include hopping between any two carbon atoms.   For pure graphene, the Hamiltonian can be written in terms of the creation and annihilation operators for electrons of spin $\sigma$ on a site in the $n^{th}$ primitive cell of the $\alpha$ and $\beta$ sublattices.   We denote the creation (annihilation) operators for the $\alpha$ and $\beta$ sublattices by  $a_{n\sigma}^\dag$  ($a_{n\sigma}$) and $b_{n\sigma}^\dag$ ($b_{n\sigma}$).  The corresponding tight-binding Hamiltonian $H_0$ for graphene may be written in real space as
\begin{widetext}
\begin{equation}
H_0 = -\sum_{ n,\delta,\sigma}(t_{\alpha \beta,\delta}a_{n,\sigma}^\dag b_{n+\delta,\sigma} + h.\ c.\ ) -
\sum_{n,\delta\neq 0,\sigma}t_{\alpha\alpha,\delta}(a^\dag_{n,\sigma}a_{n+\delta,\sigma}+b^\dag_{n,\sigma} b_{n+\delta,\sigma}).
\label{eq:h0real}
\end{equation}
Here $t_{\alpha\beta,\delta}$ and $t_{\alpha\alpha,\delta}$ ($t_{\alpha\beta},t_{\alpha\alpha}>0$) are hopping integrals which take an electron of spin $\sigma$ ($\sigma = \pm 1/2$) from a lattice site to a neighboring lattice site, and $\delta$ denotes a Bravais lattice vector of the triangular lattice.  The first sum represents hopping between sublattices and therefore all possible Bravais lattice vectors are summed over, whereas the second sum only includes hopping between sites on the same sublattice and thus, $\delta =0$ is not allowed.  We have assumed that the hopping integrals for hopping on the same sublattice are identical for the $\alpha$ and $\beta$ sublattices and therefore set $t_{\alpha\alpha}=t_{\beta\beta}$.

Eq. (\ref{eq:h0real}) can be Fourier transformed as
\begin{eqnarray}\label{eq:h0kspace}
H_0 & = & \sum_{{\bf k},\sigma}H_{0,{\bf k},\sigma}; \nonumber \\
H_{0,{\bf k},\sigma} &=& -t_{\alpha\beta}({\bf k})a^\dag_{{\bf k},\sigma}b_{{\bf k},\sigma}-t^*_{\alpha\beta}({\bf k}) a_{{\bf k},\sigma}b^\dag_{{\bf k},\sigma}-t_{\alpha\alpha}({\bf k})[a^\dag_{{\bf k},\sigma}a_{{\bf k},\sigma} + b^\dag_{{\bf k},\sigma}b_{{\bf k},\sigma}],
\end{eqnarray}
\end{widetext}

where $t_{\alpha\beta}({\bf k})= \sum_{\bf \delta}e^{i{\bf k}\cdot{\bf \delta}}t_{\alpha\beta,{\bf \delta}}$ and
$t_{\alpha\alpha}({\bf k})=\sum_{{\bf \delta}\neq 0}e^{i{\bf k}\cdot{\bf \delta}}t_{\alpha\alpha,{\bf \delta}}$.
In the limit of only nearest neighbor hopping, $t_{\alpha\alpha,\delta} = 0$,  and $t_{\alpha\beta,\delta} = 0$ except for the three nearest neighbors.  In this limit, we write $t_{\alpha\beta}({\bf k}) = t({\bf k})$.  The three nearest neighbor vectors are shown in Fig.\ \ref{fig:symmetry_points}, where they are denoted by ${\bf d}_1$, ${\bf d}_2$, and ${\bf d}_3$.  
In this limit,
\begin{equation}\label{eq:CCham}
H_{0,{\bf k},\sigma} =  -t({\bf k}) a^\dagger_{{\bf k},\sigma} b_{{\bf k,\sigma}}-t^*({\bf k})a_{{\bf k},\sigma}b^\dagger_{{\bf k},\sigma}.
\end{equation}
Here the operator $a_{{\bf k},\sigma}^\dag = \frac{1}{\sqrt{N}}\sum_n a_{n,\sigma}\exp(i{\bf k}\cdot{\bf \delta}_n)$, where $N$ is the number of primitive cells in the graphene lattice and ${\bf \delta}_n$ is the n$^{th}$ Bravais lattice vector of the triangular lattice; an analogous definition holds for $b_{{\bf k},\sigma}^\dag$.  The sum  over ${\bf k}$ is confined to the first Brillouin zone and  $t({\bf k})$ is given by \cite{Rakhmanov12}
\begin{equation}\label{eq: tk_energy}
t({\bf k})=t\left[ 1+ 2\exp\left(\frac{3 i k_x a_0}{2}\right) cos\left(\frac{\sqrt{3} k_y a_0}{2}\right)\right].
\end{equation}
 In Eq.\ (\ref{eq: tk_energy}), $t$  is the hopping energy between nearest neighbor carbon atoms ($t = 2.8 \ eV$ for graphene\cite{Rakhmanov12}), and $a_0= 1.42 \AA$ is the nearest-neighbor bond length\cite{Yazyev2007}.  

\begin{figure}[ht]
  \centering
    \includegraphics[width=0.5\textwidth]{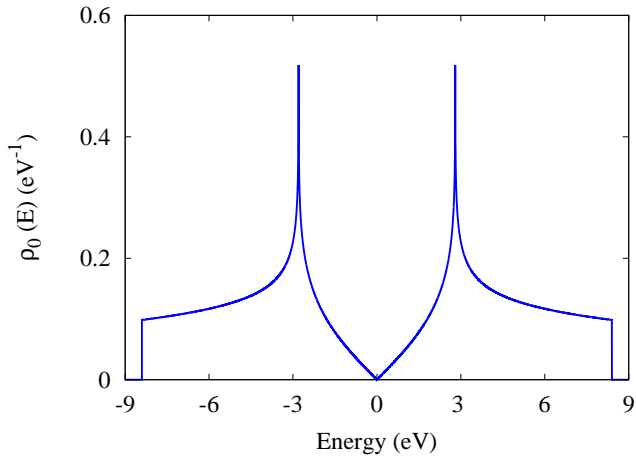}
  \caption{(Color Online) Density of states per spin and per primitive cell $\rho_{0}(E)$  for the tight-binding model defined by Eqs.\ (\ref{eq:CCham}) and (\ref{eq: tk_energy}) for the $p_z$ orbitals of graphene,  as calculated from (\ref{eq:analytic_dos})\cite{Hobson1953,Yuan2010}. Following Ref.\ \cite{Rakhmanov12}, we assume $t = 2.8 \ eV$ for graphene.}
\label{fig:analytic}
\end{figure}

We wish to investigate what happens to the density of states when when an isolated adatom is adsorbed onto the host graphene at a $T$ site\cite{nakada,chan}.  The extra piece of the tight-binding Hamiltonian, $H_I$, due to the adatom may be written in real space as
\begin{equation}
H_I = \epsilon_0 \sum_\sigma h_{0,\sigma}^\dag h_{0,\sigma} - t^\prime\sum_\sigma \left(h_{0,\sigma}^\dag a_{0,\sigma} + h_{0,\sigma}a_{0,\sigma}^\dag\right).
\label{eq:hireal}
\end{equation}
Here $h_{0,\sigma}^\dag$ and $h_{0,\sigma}$ are creation and annihilation operators for an electron of spin $\sigma$ ($\sigma = \pm 1/2$) at the site of the adatom, which we assume is located at the site $0$ of the $\alpha$ sublattice, $\epsilon_0$ is the on-site energy of an electron on that site (relative to the Dirac point of the pure graphene band structure), and $t^\prime$, ($t^\prime>0$), is the energy for an electron to hop between the adatom and the carbon atom at the site $0$ of the $\alpha$ sub-lattice.   

We now wish to express $H_I$ in terms of Bloch eigenstates of $H_0$.  These eigenstates may be written as two-component column vectors with components $\psi_1({\bf k})$ and $\psi_2({\bf k})$ satisfying the eigenvalue equation (hereafter we suppress the spin subscript until needed) 
\begin{equation}
\begin{pmatrix}
\epsilon_{\bf k}+t_{\alpha\alpha}({\bf k})  &  t_{\alpha\beta}({\bf k}) \\  t^*_{\alpha\beta}({\bf k}) &  \epsilon_{\bf k}+t_{\alpha\alpha}({\bf k}) 
\end{pmatrix}
\begin{pmatrix}
\psi_{1,{\bf k}} \\ \psi_{2,{\bf k}}
\end{pmatrix}
=0.
\vspace{0.0 in} 
\end{equation}

 The solution to this eigenvalue problem is 
\begin{equation}\label{eq:energy_graphene}
\epsilon_{\bf k} = -t_{\alpha\alpha}({\bf k})\pm|t_{\alpha\beta}({\bf k})|, 
\end{equation}
which gives the tight-binding band structure of pure graphene\cite{Rakhmanov12}, and the corresponding eigenvectors satisfy 
\begin{equation}
\psi_{1,{\bf k}} = \mp e^{-i\phi_{\bf k}}\psi_{2,{\bf k}},
\end{equation}
where the phase factor $e^{-i\phi_{\bf k}}$ is given by
\begin{equation}\label{eqn:phase}
e^{-i\phi_{\bf k}} = \frac{t_{\alpha\beta}({\bf k})}{|t_{\alpha\beta}({\bf k})|}.
\end{equation}

We can then write the destruction operator for a Bloch electron in the upper band as
\begin{equation}
\gamma_{{\bf k},1} = \frac{1}{\sqrt{2}}\left(e^{i\phi_{\bf k}}a_{\bf k} + b_{\bf k}\right),
\end{equation}
and in the lower band as
\begin{equation}
\gamma_{{\bf k},2} = \frac{1}{\sqrt{2}}\left(e^{i\phi_{\bf k}}a_{\bf k} - b_{\bf k}\right),
\end{equation}
where we have defined $\gamma_{{\bf k},1}$ and $\gamma_{{\bf k},2}$ to be properly normalized, so that, for example, the anticommutator 
$\{\gamma_{{\bf k},1}^{\hspace{1pt }}  , \gamma^\dag_{{\bf k},1}\} = 1$. 

With these definitions, we can now use the inverse Fourier transform of the $\gamma_{{\bf k},1}$ and $\gamma_{{\bf k},2}$ to obtain
\begin{equation}
a_n = \frac{1}{\sqrt{2N}}\sum_{\bf k}e^{-i{\bf k}\cdot{\bf \delta}_n}e^{-i\phi_{\bf k}}(\gamma_{{\bf k},1} + \gamma_{{\bf  k},2}).
\end{equation}
Thus, we can rewrite the impurity Hamiltonian (\ref{eq:hireal}) in momentum space as
\begin{widetext}
\begin{equation}
H_I = \epsilon_0 h_0^\dag h_0 - \frac{t^\prime}{\sqrt{2N}} \left[ h_0^\dag \sum_{\bf k}e^{-i\phi_{\bf k}}(\gamma_{{\bf k},1} +\gamma_{{\bf k},2}) + h.c.\right].
\label{himp}
\end{equation} 
Thus, in $H_I$,  the creation and annihilation operators of the adatom are connected to every eigenstate of the graphene band structure by matrix elements of equal magnitude (though different phase). For a hydrogen adatom,  we take $\epsilon_0 = 0.4 \ eV$ and $t' = 5.8
\ eV$, as found in \cite{Rakhmanov12}.  The one-electron Hamiltonian, $H_0 + H_I$, is a special case  of the Anderson impurity model\cite{anderson1961}, where the impurity state is coupled to all the band electron states by matrix elements of equal magnitude.

We also include in our calculation the effects of an on-site electron-electron interaction of the Hubbard form,
\begin{equation}\label{hubbard_u}
H_U= U n_{0\uparrow} n_{0,\downarrow},
\end{equation}
where  $n_{0,\sigma} = h_{0,\sigma} ^\dag h_{0,\sigma}$ is the number of electrons with spin $\sigma$ on the hydrogen site.  For a hydrogen adatom we take $U$ to be the difference between the ionization potential and the electron infinity providing us with a numberical value of $U \sim 12.85 eV = 4.59t$\cite{pariser,lykk}.

The Hubbard term given in Eq. (\ref{hubbard_u}) is quartic in the creation and annihilation operators. Therefore, in order to calculate the properties of the Hamiltonian including this term, we use a standard mean field theory to  rewrite this term (see, e.\ g., Ref.\ \cite{Fazekas1999}) in the form
\begin{equation}\label{MFT_U}
H_U  \sim U \left[h_{0\uparrow}^\dagger h_{0\uparrow} \langle n_{0\downarrow}\rangle+h_{0\downarrow}^\dagger h_{0\downarrow} \langle n_{0\uparrow}\rangle- \langle n_{0\uparrow}\rangle \langle n_{0\downarrow}\rangle\right].
\end{equation}
\end{widetext}
With this approximation, the total Hamiltonian, consisting of the sum of Eqs.\ (\ref{eq:CCham}), (\ref{himp}), and (\ref{MFT_U}), becomes quadratic in electron creation and annihilation operators, and can be diagonalized.  The Fermi energy, total energy, and magnetic properties of the system can then be obtained by an iterative process as described below.  The electronic density of states corresponding to the one-electron Hamiltonian, $H_0+H_I$, can be obtained analytically, as we describe below, which makes the calculation of the total energy and the magnetic properties quite simple. 
 
\section{Green's function, Density of States, and Spectral Function of Graphene-Adatom System}
\subsection{Green's Function}

We use the single particle Green's function approach to calculate the local and total density of states of the graphene-adatom system, initially omitting the Hubbard-U term.  We continue to  suppress the spin degree of freedom since, in the absence of the Hubbard term, spin just gives an extra factor of 2.   To that end, we first introduce the resolvent operator
\begin{equation}
G(z) = \frac{1}{z - H},
\label{eq:resolvent}
\end{equation}
where $z = E + i\eta,$ ($\eta \rightarrow 0^+$), and $H = H_0 + H_I$.   If there are $2N$ carbon atoms and 1 adatom, $G(z)$ can be expressed as an $(2N+1)\times (2N+1)$-dimensional matrix.    It is convenient to use the 2N Bloch states (corresponding to $N  {\bf k}$ values) created by the operators $\gamma_{{\bf k},1}^\dag$ and $\gamma_{{\bf k},2}^\dag$ as the basis for this matrix, plus the adatom orbital corresponding to $h_0^\dag$.   If we let the adatom orbital correspond to the first of the $(2N+1)$ states, then one can easily write out the matrix $z-H$ of which $G(z)$ is the inverse.

\subsection{Density of States}

We denote the local electronic density of states per spin on the adatom site by $\rho_{00}(E)$.    $\rho_{00}(E)$ is related to $G(z)$ by
\begin{equation}
\rho_{00}(E) = -\frac{1}{\pi}\mathrm{Im} G_{00}(z) = -\frac{1}{\pi}\mathrm{Im}\langle 0|\frac{1}{z - H}|0\rangle.
\label{eq:ldos}
\end{equation}
Here $z = E +i\eta$, ($\eta \rightarrow 0^+$), and $\langle 0|1/(z-H)|0\rangle$ denotes the matrix element of $1/(z-H)$ evaluated at the location of the adatom, which we take to be above the atom $0$ on the $\alpha$ sub-lattice.  This  corresponds to an element in the first row and first column of the matrix $(z-H)^{-1}$.  We can obtain this matrix element as
\begin{equation}
G_{00}(z) = \frac{\mathrm{cof}_{00}(z-H)}{\mathrm{det}(z-H)},
\label{eq:cofdet}
\end{equation}
where cof$_{00}(z-H)$ denotes the cofactor of the element in the first row and first column of the matrix $z-H$, and the denominator is the determinant of $z-H$.  Both quantities are readily evaluated, and the result for $\rho_{00}(E)$ is
\begin{equation}\label{eq:local_dos}
\rho_{00}(E) = -\frac{1}{\pi}\mathrm{Im}\left(\frac{1}{z - \epsilon_0 - \frac{t^{\prime 2}}{2N}{\cal G}_0(z)}\right) _{z=E+i0+},
\end{equation}
where
\begin{equation}\label{eq:script_G}
{\cal G}_0(z) =\sum_{{\bf k},\lambda=1}^2\left(\frac{1}{z-\epsilon_{{\bf k},\lambda}}\right) \equiv \mathrm{Tr}\left(\frac{1}{z - H_0}\right)
\end{equation}
and $\epsilon_{{\bf k}, 2} = -\epsilon_{{\bf k},1}$ is given by Eq.\ (\ref{eq:energy_graphene}).

$\mathrm{Im} \ {\cal G}_0(z)$ is related to the (unperturbed) graphene density of states per graphene unit cell (per spin), which we denote $\rho_{0}(E)$, by
\begin{equation}\label{eq:dos_analytic}
-\frac{1}{\pi}\mathrm{Im}\ {\cal G}_0(E+i\eta)=N \rho_{0}(E),
\end{equation}
with $\eta \rightarrow 0^+$ \cite{anderson1961,wehling}.  For the form of $H_0$ which includes  only nearest neighbor hopping,  $\rho_0(E)$ is\cite{Hobson1953,Yuan2010} 
\begin{numcases}{\rho_{0}(E)=}\label{eq:analytic_dos}
 \frac{2 E}{t^2 \pi^2} \frac{1}{\sqrt{f(x)}}K\left( \frac{4x }{f(x)}\right)  &  $0<x<1$   \\
\frac{2 E}{t^2 \pi^2}\frac{1}{\sqrt{4x}}K\left(\frac{f(x)}{4x}\right) &$1<x<3$, \nonumber 
\end{numcases}
where $x= E/t$, $f(x) = (1+x)^2 - \frac{(x^2-1)^2}{4}$ and $K(m)$ is the elliptic integral of the first kind. We plot Eq.  (\ref{eq:analytic_dos}) in Fig.\ (\ref{fig:analytic}) normalized such that $\int_{-3t}^0\rho_0(E)dE = 1$.

$\mathrm{Re} \ {\cal G}_0(E)$ is related to Eq.  (\ref{eq:analytic_dos}) via the principal value integral\cite{anderson1961,wehling}
\begin{equation}\label{eq:principle_real}
\mathrm{Re} \ {\cal G}_0(E) = N \ \mathrm{P}\left(\int_{-3t}^{3t}\frac{\rho_0(E')}{E-E'}dE'\right),
\end{equation}
where the integral runs over the range where $\rho_0(E^\prime) \neq 0$. The density of states on the carbon sites (per spin) in the presence of an adatom may be written as $\rho_g(E) = -\frac{1}{\pi}\mathrm{Im}\sum_{{\bf k},\lambda}\langle {\bf k},\lambda|\frac{1}{z - H}|{\bf k},\lambda\rangle$, where $z = E +i\eta$, ($\eta\rightarrow 0^+$).  Each of the elements of this sum can be computed using the analog of Eq.\ (\ref{eq:cofdet}), with the result
\begin{equation}\label{eq:change_carbon}
\rho_{g}(E) = N\rho_{0}(E) +\frac{1}{\pi}\mathrm{Im}\left(\frac{t^{\prime 2}/2N}{z-\epsilon_0-\frac{t^{\prime 2}}{2N}{\cal G}_0(z)}\frac{d{\cal G}_0}{dz}\right)_{z=E+i0^+}.
\end{equation}

\begin{widetext}

The {\it total} density of states per spin is is the sum of the expressions in Eqs.\ (\ref{eq:local_dos}) and (\ref{eq:change_carbon}), and can be rearranged to have the form
\begin{equation}\label{eq:total_dos}
\rho_{tot}(E) = \rho_g(E) + \rho_{00}(E) = N\rho_{0}(E) -\frac{1}{\pi}\mathrm{Im}\left( \frac{d}{dz}\mathrm{ln}[z-\epsilon_0-\frac{t^{\prime 2}}{2N}{\cal G}_0(z)]\right)_{z=E+i0^+}.
\end{equation}

\subsection{Spectral Function}

We can use an analogous approach to calculate the spectral function $A({\bf k}, E)$.   $A({\bf k}, E)$ represents the probability density that an electron  with Bloch wave-vector ${\bf k}$  has energy $E$, and is given by
\begin{equation}
A({\bf k}, E) = -\frac{1}{\pi}\mathrm{Im}\sum_\lambda\langle {\bf k}, \lambda|\left(\frac{1}{z-H}\right)|{\bf k}, \lambda \rangle.
\end{equation}
These matrix elements can be evaluated using the methods of the preceding section, with the result

\begin{equation}
A({\bf k}, E) = -\frac{1}{\pi}\mathrm{Im}\sum_{\lambda=1}^2\left[\frac{1}{z-\epsilon_{{\bf k},\lambda}} +
\frac{t^{\prime 2}}{2N}\left(\frac{1}{z-\epsilon_0 - \frac{t^{\prime 2}}{2N}{\cal G}_0(z)}\right)\frac{1}{(z-\epsilon_{{\bf k},\lambda})^2}\right]_{z=E+i0^+}.
\label{eq:spectral}
\end{equation}

It is convenient to write this spectral function in terms of a self-energy $\Sigma_\lambda({\bf k}, E)$ as
\begin{equation}
A({\bf k}, E) = -\frac{1}{\pi}\mathrm{Im}\left(\sum_{\lambda=1}^2\frac{1}{z - \epsilon_{{\bf k},\lambda} - \Sigma_\lambda({\bf k}, z)}\right)_{z=E+i0^+},
\label{eq:selfe}
\end{equation}
where $\Sigma_\lambda({\bf k}, E)$ is readily computed by equating Eqs. (\ref{eq:spectral}) and (\ref{eq:selfe}).  To first order in $1/N$, $\Sigma_\lambda({\bf k}, z)$ is found to be independent of both $\lambda$ and ${\bf k}$ and to take the form
\begin{equation}
\Sigma_\lambda({\bf k}, z) =  \frac{t^{\prime 2}}{2N}\left(\frac{1}{z-\epsilon_0-\frac{t^{\prime 2}}{2N}{\cal G}_0(z)}\right)_{z=E+i0^+}.
\label{eq:selfe1}
\end{equation}

All these equations [(\ref{eq:local_dos}), (\ref{eq:script_G}), (\ref{eq:dos_analytic}), (\ref{eq:principle_real})-(\ref{eq:total_dos}), (\ref{eq:selfe}), and (\ref{eq:selfe1})] remain valid for non-nearest-neighbor hopping; only the form of the graphene density of states has to be changed.
\end{widetext}

\subsection{Effects of Electron-Electron Interaction; Spin-Polarized Density of States and Magnetic Moment}\label{sectionD}

Finally, we discuss the effects of including a non-zero Hubbard term  $H_U$ [Eq.\ (\ref{hubbard_u})] in the Hamiltonian.   If we treat $H_U$ by mean-field theory [Eq.\ (\ref{MFT_U})], then the densities of states for spin-up and spin-down electrons may be different.  We can calculate these partial densities of states  self-consistently as follows.   First,  we make an initial assumption for the value of $\langle n_{0\uparrow}\rangle$ and $\langle n_{0\downarrow}\rangle$. Then the effective on- site energy for an up-spin electron on the hydrogen adatom is obtained by making the replacement 
\begin{equation}\label{eq:includespin}
\epsilon_{0,\uparrow} \rightarrow \epsilon_0 + U\langle n_\downarrow\rangle ,
\end{equation}
with a corresponding expression for $\epsilon_{0,\downarrow}$.   Given $\epsilon_{0,\uparrow}$ and $\epsilon_{0,\downarrow}$, we can compute the local  densities of states $\rho_{00,\uparrow}$ and $\rho_{00,\downarrow}$ using the appropriate generalizations of Eq.\ (\ref{eq:local_dos}); we can also obtain the total densities of states $\rho_{tot,\uparrow}$ and $\rho_{tot,\downarrow}$ using the corresponding generalizations of Eq. (\ref{eq:total_dos}).    The Fermi energy, $E_F$, is then obtained from the condition
\begin{equation}\label{number_dos}
2N+1= \int_{-3t}^{E_F}\rho_{tot}(E) dE,
\end{equation}
where we assume one adatom, $2N$ carbon sites, and $\rho_{tot}(E) =\rho_{tot,\uparrow}(E) + \rho_{tot,\downarrow}(E)$.  Given $E_F$, we then recalculate $\langle n_{0,\uparrow}\rangle$ and $\langle n_{0,\downarrow}\rangle$.   The procedure is repeated until successive iterations do not lead to a significant change in $\langle n_{0,\uparrow}\rangle$ and $\langle n_{0,\downarrow}\rangle$.   In practice, we require that these quantities change by no more than $\pm 0.001 n_a$
on successive iterations (here $n_a$ is the number of adatoms in the calculation). Typically, about twenty iterations  of the self-consistent equations are needed to attain this degree of convergence, as discussed further below. 

Once $E_F$ has been found, the total magnetic moment $\mu_T$ of the system is obtained from
\begin{equation}
\label{eq:magmom}
\frac{\mu_T}{\mu_B}  =  \int_{-3t}^{E_F}\left[\rho_{tot,\uparrow}(E)-\rho_{tot,\downarrow}(E)\right]dE,
\end{equation}
where $\mu_B$ is the Bohr magneton.

In the limit $U \rightarrow \infty$, the mean-field version of the Hubbard model can be done without iteration.   In this limit, only one of the quantities $\langle n_\uparrow\rangle$ or $\langle n_\downarrow \rangle$ is non-zero.   The reason is that if, say, $\langle n_\uparrow\rangle$ is non-zero, then the energy to put a spin-down electron on the adatom becomes infinite, and hence the number of spin-down electrons must be zero.   To be definite, we assume that $\langle n_\downarrow \rangle = 0$.  In that case, we just have $\epsilon_{0,\uparrow} = \epsilon_0$, and $\epsilon_{0,\downarrow} \rightarrow \infty$.  The total density of states for the up spins will then be given by Eq. (\ref{eq:total_dos}), while that for the down spins is just that of unperturbed graphene: $\rho_{tot,\downarrow} = N\rho_0(E)$. 

 In the limit $U\rightarrow \infty$ the Fermi energy, $E_F$, is obtained from Eq. (\ref{number_dos}) and  may be simplified to
\begin{widetext}
\begin{equation}\label{eq:extra_adatom}
\int_0^{E_F} 2N\rho_0(E)dE  -   
\frac{1}{\pi}\mathrm{Im}\ln\left[\frac{E_F -\epsilon_0-\frac{t^{\prime2}}{2N}{\cal G}_0(E_F)}{-3t-\epsilon_0-\frac{t^{\prime2}}{2N}{\cal G}_0(-3t)}\right]  =  1.
\end{equation}
\end{widetext}
Once $E_F$ has been obtained, the magnetic moment $\mu_T$ can be again found using Eq.\ (\ref{eq:magmom}).     Since both $\rho_\uparrow(E)$ and $\rho_\downarrow(E)$ are available analytically, using Eqs. (\ref{eq:magmom}) and (\ref{eq:extra_adatom}), $\mu_T$ is easily computed in closed form.

\section{Numerical Results}

\begin{widetext}

\begin{figure}[h t]
  \centering
    \includegraphics[width =0.9 \textwidth]{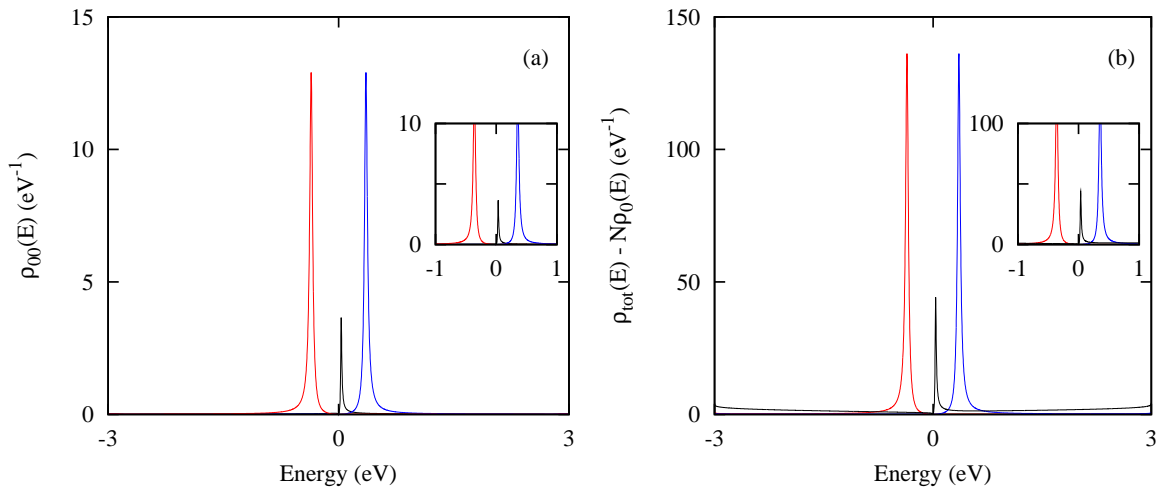}
  \caption{(Color Online) (a).  Local density of states per spin on the adatom site, $\rho_{00}$(E),  with $U=0$  for a graphene sample with N = 500 graphene unit cells (1000 C atoms) and one adatom.  We assume the model described in the text [Eqs.\ (\ref{eq:CCham}), (\ref{eq: tk_energy}), and (\ref{himp})].     Black curve: $t^\prime = 5.8  \ eV$, $\epsilon_0 = 0.4 \  eV$;
blue curve: t$^\prime =  1.0 \ eV$, $\epsilon_0 = 0.4 \ eV$; 
and red curve: t$^\prime = 1.0 \ eV$, $\epsilon_0 = -0.4 \ eV$.   In all three cases, $t = 2.8 \  eV$.   The Fermi energy is calculated using Eq.\ (\ref{number_dos}) and gives $E_F = 0.236 \ eV$.
(b). The change in the density of states per spin due to the adatom for the three cases shown in (a).  In both (a) and (b), the insets are enlargements of the region between -1 eV and +1 eV.}
\label{fig:ldos_hydrogen}
\end{figure}

\end{widetext}

 In Fig.\  \ref{fig:ldos_hydrogen}(a), we plot the local density of states $\rho_{00}(E)$ for parameters appropriate to a hydrogen adatom on graphene with $U = 0$, as calculated from Eq.\ (\ref{eq:local_dos}).   We use the parameters $t = 2.8 \ eV$, $t^\prime = 5.8 \ eV$, and $\epsilon_0 = 0.4\ eV$, as given by Ref.\  \cite{Rakhmanov12} for a hydrogen adatom.  In Fig.\ \ref{fig:ldos_hydrogen}(b), we plot the {\it change} in the total density of states produced by a single hydrogen atom, i.\ e.,  the quantity $\rho_{tot}(E) - N\rho_0(E)$ for the three cases of Fig.\ \ref{fig:ldos_hydrogen}(a), calculated using Eqs. (\ref{eq:ldos}) and (\ref{eq:total_dos}).  

\begin{figure}
\centering
\includegraphics[width = 0.5 \textwidth]{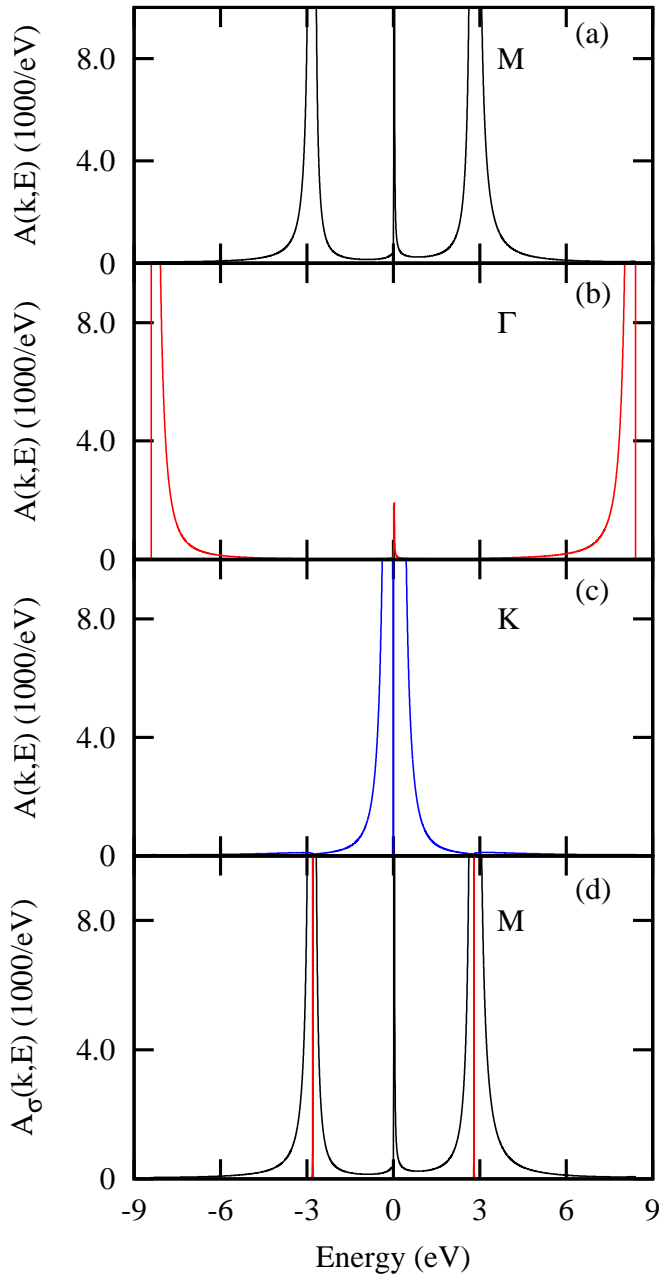}
\caption{(Color Online) (a)-(c). Spectral function $A({\bf k}, E)$ at $U=0$ for the graphene-adatom system, as calculated  using Eqs.\ (\ref{eq:selfe}) and (\ref{eq:selfe1}) for three values of ${\bf k}$ corresponding to $M$, $K$, and $\Gamma$ respectively, assuming a nearest-neighbor tight-binding band.    We use  $t=2.8 \ eV$ , $t'=5.8 \ eV$, $\epsilon_0 = 0.4 \ eV$, and $N = 500$.  
(d). The spin polarized spectral function for the point $M$ and the case $U \rightarrow \infty$ in the mean-field approximation, using the same parameters as in (a)-(c).   $A_{\sigma}({\bf k }, E)$ for the majority spin component is shown in black and that of the minority component in red.  For the minority spin component, $A_{\sigma}({\bf k}, E)$ is just that of the unperturbed graphene sheet, i.\ e., delta functions at the unperturbed pure graphene energy eigenvalues for all values of ${\bf k}$.   For the case $U = 0$, and  for the majority spin at $U = \infty$, the adatom contribution occurs near $E_F$, while the graphene sheet contribution corresponds to broadened peaks near the unperturbed energy eigenvalues $\epsilon_{\bf k}$ given in Eq. (\ref{eq:energy_graphene}).  For points $\Gamma$ and $M$, the integral of the peak near $E \sim 0$ is of order 1/(2N) times those of the main peaks in the spectral function, and thus vanishes as $N \rightarrow \infty$. }
\label{fig:spectral}
\end{figure}

Next, we calculate both the spectral function $A({\bf k}, E)$ for $U = 0$ and the spin polarized spectral function $A_\sigma({\bf k}, E)$  for $U = \infty$ as functions of $E$ for several values of ${\bf k}$, using the parameters of Fig.\ \ref{fig:analytic}.    $A_\sigma({\bf k}, E)$ is obtained using a generalization of  Eq.\ (\ref{eq:selfe}) in the limit $U \rightarrow \infty$ as discussed in subsection \ref{sectionD}.  The resulting spectral functions are shown in Fig.\ \ref{fig:spectral} through first order in $1/N$.  The contribution from the adatom appears as the sharp spike near $E_F=0.173 \ eV$, while the contribution from the graphene sheet is shown as broadened peaks near the values of $\epsilon_{{\bf k},i}$ $(i = 1, 2)$ for the three choices of ${\bf k}$.   The self energy term [Eq.\ (\ref{eq:selfe1})]  controls the width of the graphene resonances.   The integral of the graphene sheet's contribution to the spectral function will be of order $N$ times larger then that of the adatom.  Furthermore, the width of the graphene peaks in the spectral function is proportional to the density of adatoms.

\begin{widetext}

\begin{figure}[h t]
  \centering
    \includegraphics[width=1.0\textwidth]{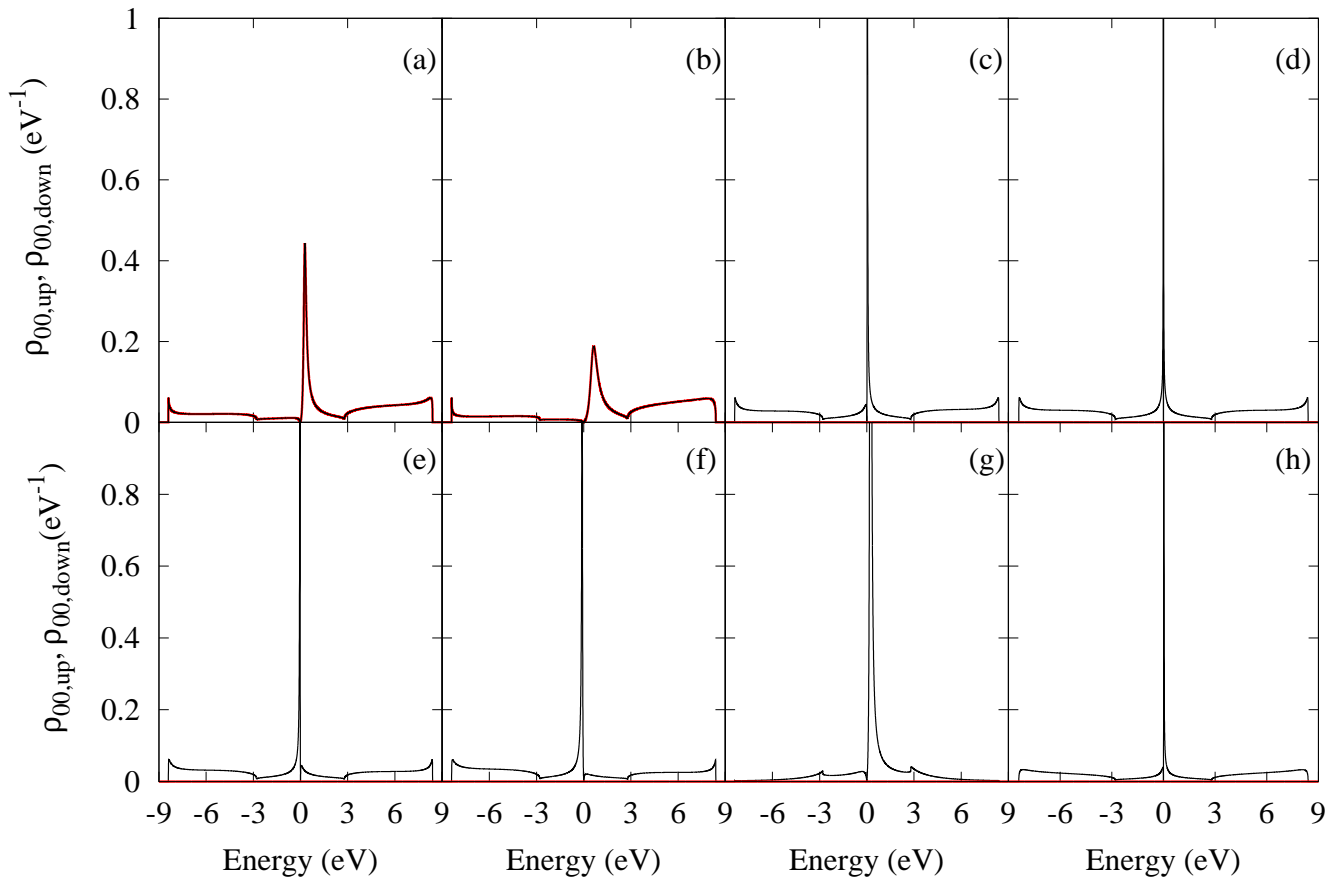}
  \caption{(Color Online) Spin-polarized local density of states (sLDOS) per adatom for both the majority spin (black solid line) and  minority spin (red solid line), as obtained by substituting Eq.\ (\ref{eq:includespin}) into  Eq.\ (\ref{eq:local_dos}).   We use $t = 2.8 \ eV$ and the following 
values of the on-site energy $\epsilon_0$, hopping energy $t^\prime$, and Hubbard energy $U$:
(a). $(t',\epsilon_0,U) = (5.8,0.4,4.59 t)$; (b). $(5.8,0.4,10t)$; (c)$(5.8,0.4,\infty)$ (d). $(5.8,0.0,\infty)$; (e). $(5.8,-0.4,\infty)$; (f). $(5.8,-1.0,\infty)$; (g). $(1.8,0.4,\infty)$; (h). $(7.8,0.4,\infty)$.}
\label{fig:changeparameter}
\end{figure}

\end{widetext}


\begin{figure}[hbt]
\centering
\includegraphics[width=0.5\textwidth]{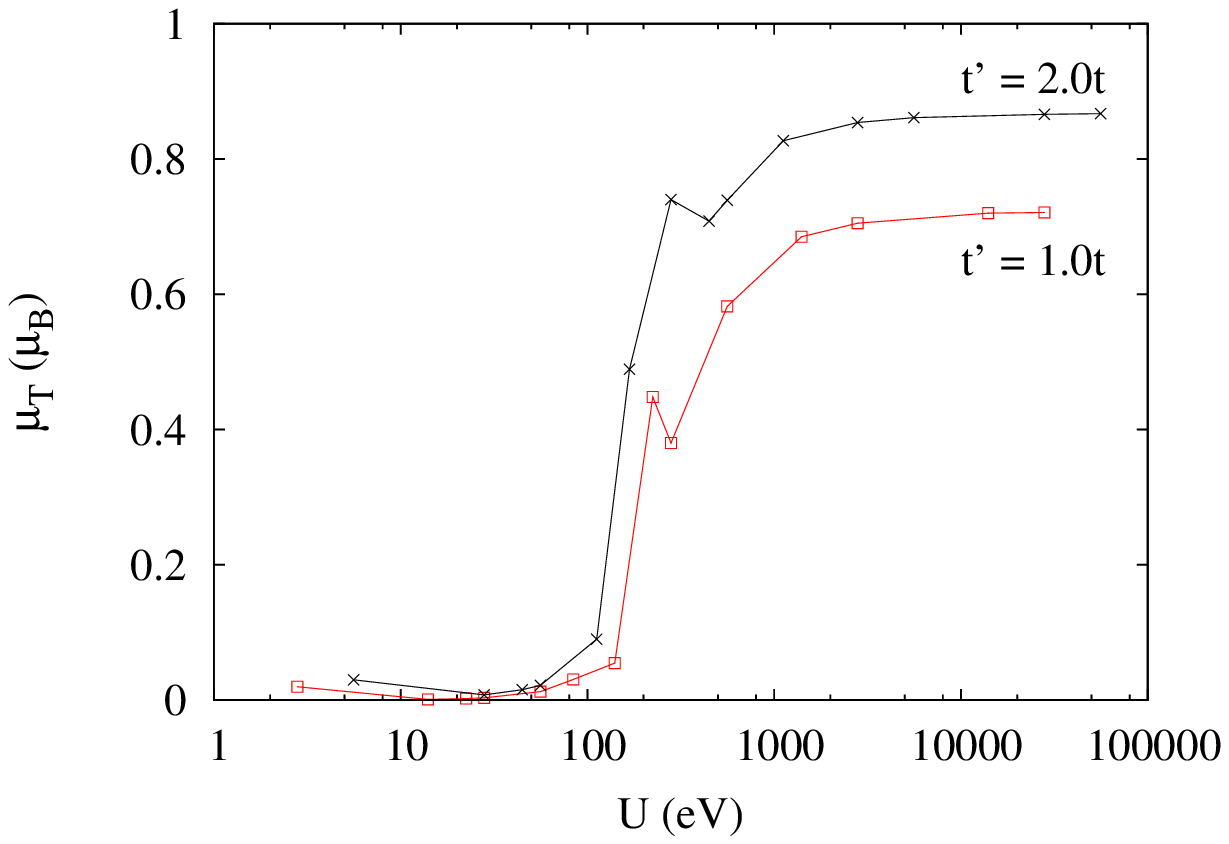}
\caption{Total magnetic moment on the adatom and the graphene ($\mu_T$), in units of $\mu_B$, versus $U$ for two values of $t^\prime/t$ (Red curve: $t^\prime = 1.0t$, Black curve $t^\prime = 2.0t$).  In this plot, we hold $\epsilon_0 = 0.4 eV $ and $t= 2.8 eV$  as we vary $U$ for two different choices of $t^\prime$.    In both cases, the moment seems to become nonzero at a characteristic value of $U$,  which depends on $t^\prime$, $t$, and $\epsilon_0$.  Lines connect calculated points.}
\label{fig:magmom4}
\end{figure}

Using the mean-field methods described in section \ref{model} we can calculate a variety of other spin-independent and spin-dependent properties of the adatom-graphene system. These include $\rho_{00,\uparrow}(E)$ and $\rho_{00,\downarrow}(E)$, the local density of states of up and down spin on the adatom; the induced magnetic moment on the adatom  ($\mu_a$) and in the entire system of graphene sheet plus adatom ($\mu_T$); and the net charge transfer from the adatom to the sheet, all as functions of the parameters $U$, $\epsilon_0$, and $t^\prime$.  The magnetic moment on the adatom site is  $\mu_a = (\langle n_\uparrow\rangle - \langle n_\downarrow\rangle)\mu_B$.  $\langle n_\sigma\rangle$ is obtained from 
\begin{equation}
\langle n_\sigma \rangle = \int_{-\infty}^{E_F} \rho_{00,\sigma}(E) dE,
\end{equation}
where $\rho_{00,\sigma}$ ($\sigma = \uparrow or \downarrow$) is defined using the appropriate generalization of Eq.\ (\ref{eq:local_dos}).  The total magnetic moment is given in Eq. (\ref{eq:magmom}). The net charge transfer from adatom to the graphene lattice is obtained by first integrating $\rho_{00,\uparrow}+\rho_{00,\downarrow}$  up to the Fermi energy, to obtain the net number of electrons on the adatom, then subtracting this quantity from the adatom valence $Z$  (i.\ e., for hydrogen, Z = 1) to obtain the net charge transfer.   

We have carried out these calculations for various values of the adatom on-site energy $\epsilon_0$, Hubbard parameter $U$, and hopping energy $t^\prime$.  In Table I we summarize the results above for parameters appropriate to a hydrogen adatom and summarize the trends when the various adatom parameters are varied. Additional results are shown in Figs.\ \ref{fig:ldos_hydrogen} and \ref{fig:changeparameter}.   As can be seen in Table I and Fig.\ \ref{fig:changeparameter}, the parameter values thought to be appropriate to an $H$ adatom ($U = 4.59t$, $t^\prime = 5.8 \ eV$, and $\epsilon_0= 0.4 \ eV$), lead to a very small magnetic moment on the adatom though there is an increase in both the LDOS and TDOS close to the Fermi energy ($E_F = 0.173 \ eV$; see Fig.\ \ref{fig:changeparameter}(b)).

\begin{table}[h t]
\begin{center}
\begin{tabular}{| c | c | c| c|c|c| c |}
\hline
\multicolumn{7}{  | c|  } {Summary of Numerical Results} \\
\hline
U & $\epsilon_0$ & $t^\prime$& $\mu_a$  ($\mu_B$)& $\mu_T $ &  Charge  & $E_F$\\
& (eV)&(eV)  & per & ($\mu_B$)  & Transfer& (eV)  \\
& & &  adatom & & ($|e|$)  &  \\
\hline\hline
$0.0t ^a $ & $ 0.4 $ & 5.8 & $0.0$ &  0.0  & $  0.372$ &0.372\\   
\hline
$ 4.59 t^b $ & $0.4 $& 5.8 & 2.67E-4  & 1.13E-3  & 0.695& 0.173\\   
\hline
$ 10.0 t$& 0.4& 5.8&  3.11E-3&7.67E-3 & 0.758  &0.236 \\    
\hline
$ \infty $ & $0.4 $ & $5.8$ &  0.300 &0.871 &  0.699 &0.111\\   
\hline
$\infty $  & $ 0.0 $ & 5.8 & 0.260 & 0.927  & 0.738 & 0.050 \\  
\hline 
 $\infty$ & $-0.4$ & 5.8& 0.338  &0.958 &0.662 &0.050 \\  
\hline
$\infty$ & $-1.0 $& 5.8  &0.360  &0.990 & 0.639 &-0.01  \\ 
\hline
$\infty$ & 0.4 &1.8 &0.358  & 0.506  &0.641 & 0.236 \\ 
\hline
$\infty$& 0.4& 7.8&  0.219& -0.11 &0.780 &0.236 \\ 
\hline
\end{tabular}
\caption{Magnetic moment on the adatom ($\mu_a$), total magnetic moment on the graphene-adatom system ($\mu_T$) (both in units of $\mu_B$), and charge transferred from the adatom to the graphene lattice (in units of an electron charge), for various choices of $U$ , on-site energy $\epsilon_0$, and hopping energy $t^\prime $.   Note that $U=4.59 t$, $\epsilon_0=0.4 \ eV$, and $t^\prime=5.8 \ eV$  corresponds to the expected parameters of a hydrogen adatom.  When $U \rightarrow \infty$, we find a spin polarized state near the Fermi energy.  The magnetic moment calculated on the adatom is done using a combination of  Eqs.  (\ref{MFT_U}) and (\ref{eq:local_dos}) and the magnetic moment on the sheet is calculated using Eq. (\ref{eq:magmom}).
 \\ (a)-  Spin polarized LDOS  plotted in Fig. (\ref{fig:ldos_hydrogen}). \\(b)-  Spin polarized LDOS plotted in Fig. (\ref{fig:changeparameter}b).}
\end{center}
\end{table}

In general, as seen in Table I, for sufficiently large $U$, a nonzero magnetic moment develops on both the adatom and the graphene sheet.  The moment on the adatom is of order $0.3 \ \mu_B$ in this limit, for the given parameters, while the sum of the moments on the adatom and the sheet approaches $\mu_B$ in this limit.   We also find in all of our calculations that a large fraction (typically $0.6-0.7$ of the electron) is transferred from the adatom to the graphene sheet for the parameters we consider. 
For the parameters appropriate to hydrogen adatoms, the model predicts no, or only a very small, induced magnetic moment.   A possible explanation is that our model assumes no lattice distortion due to the adatom.  But DFT calculations have shown that the surface of graphene is warped due to the addition of an adatom.  This warping could change the distance between the adatom and the neighboring carbon atoms, and hence possibly the value of the Coulomb integral.

In Fig.\ \ref{fig:magmom4}, we show the total magnetic moment of the system as given by Eq.\ (\ref{eq:magmom}) plotted as a function of $U$, for various values of $t^\prime$. In each case shown, $\epsilon_0 = 0.4 \ eV$ and $t = 2.8 \ eV$.   In all the plots, there is an apparent threshold behavior: the moment becomes nonzero only if $U$ exceeds a threshold value which depends on $t^\prime$ as well as on $U$.   While these calculations are done using a simple mean-field approximation, they seem to be consistent with other work on related models\cite{anderson1961,fradkin}.

\section{Discussion}

Using a tight-binding model we have calculated the local and total density of states and the spectral function for a system consisting of a single adatom in a $T$ site on graphene.  Because the hopping integral from the adatom to a graphene Bloch state has the same magnitude for any ${\bf k}$, we have shown that these quantities can be calculated analytically.   This simplification holds even if we do not make the oft-used linear approximation\cite{wehling} for the graphene density of states near the Dirac point.  It is also valid even if we include non-nearest-neighbor hopping in the tight-binding graphene Hamiltonian.  Since our numerical results give both the local and total density of states, we can compute the charge transfer from the adatom to the graphene.  Our numerical results show that, for most parameters we consider, this charge transfer is a substantial fraction of an electron (approximately $70 \%$  for parameters appropriate to  hydrogen).    

Because the calculations are at low adatom concentration, the adatom-induced density of states is linear in concentration.  Other work has treated the same system at higher adatom concentration, but only numerically\cite{Rakhmanov12}.  In future work, it might be possible to treat the present model analytically at higher concentrations, at least approximately.  It would also be of interest to  include effects of lattice distortions, which are known to exist when  adatoms bind to graphene \cite{boukhvalov}, and which can lead to a large increase in spin-orbit interactions \cite{neto,Gmitra}. Such spin-orbit interactions would likely have a large effect on the magnetic properties arising from the adatom. 

We have also calculated the magnetic properties induced by the adatom, using a Hubbard model treated within mean-field theory. For all choices of the Hamiltonian parameters, we find that there is a critical value of the Hubbard $U$ above which the density of states near the Fermi energy is spin-polarized and a net induced magnetic moment is formed.  The appearance of this magnetic moment was predicted long ago to occur within mean-field theory for models with a slowly
varying density of states near the Fermi energy\cite{anderson1961}.   Here, it is also found to occur in a system with a roughly linear density of states near $E_F$. While the mean-field approximation is probably unreliable for this model, we note that a similar threshold for moment formation was also found, within the Kondo Hamiltonian, for a system with a linear density of states\cite{fradkin}.  Since, at large $U$, it is known that the Hubbard model can be approximately transformed into the Kondo model\cite{schrieffer}, it seems plausible that there could really be a threshold behavior in the Hubbard model with a linear density of state  such as is found here, even though we use a mean field theory to obtain it. 

A somewhat counterintuitive result of our calculations is that, as $t^\prime$ increases, the value of $U$ needed to induce a magnetic moment becomes smaller.    Since a larger $t^\prime$ suggests that it is easier for the electron to hop from the impurity to the graphene, one might expect that a moment on the impurity atom would be less likely to form.   A possible explanation is that the larger $t^\prime$ also causes the peak in the impurity density of states to shift closer to the Dirac point, where the graphene density of states is smaller.  Thus, there are fewer final states available for an electron to hop into, and hence, the electron is less likely to hop, thus increasing the likelihood of moment formation on the impurity.    

In our approximation we also calculate the spectral function of our graphene-adatom system to first order in $1/N$. The main effect of the adatom is, as expected, simply to broaden the delta-function peaks that the spectral function would exhibit in an ideal graphene lattice. In our approach, this broadening, and the shape of the spectral line, are computed analytically.  To the same order, we find that both the adatom contribution to the spectral function at $E\sim0$ and the broadening of the graphene spectral lines will vanish as $N\rightarrow \infty$.   

In summary, we have, using a single-particle Green's function approach together with a tight binding Hamiltonian in the limit of no electron-electron correlations, obtained analytical equations for the LDOS, TDOS, and spectral function for adatoms on the surface of graphene.  Using the same model with a finite Hubbard energy $U$, we find that a magnetic moment is induced both on the adatom and nearby on the graphene sheet above a critical value of $U$ which depends on the other model parameters.   These results are not only of intrinsic interest but also may be useful in understanding the behavior of a variety of adatoms on graphene.    

\section{Acknowledgments}
This work was supported by the Center for Emerging Materials at The Ohio State University, an NSF MRSEC (Grant No.\ DMR0820414) .  We thank Prof.\  Roland Kawakami for valuable discussions.

\end{document}